\documentclass[aps,prb,preprint,groupedaddress,floatfix]{revtex4}
\usepackage{graphicx,amssymb}
\begin{document}

\title{Unusual crystallization behavior close to the glass transition.} 
\author{Caroline Desgranges, and Jerome Delhommelle}
\affiliation{Department of Chemistry, University of North Dakota, Grand Forks ND 58202}
\date{\today}

\begin{abstract}
Using molecular simulations, we shed light on the mechanism underlying crystal nucleation in metal alloys and unravel the interplay between crystal nucleation and glass transition, as the conditions of crystallization lie close to this transition. While decreasing the temperature of crystallization usually results in a lower free energy barrier, we find an unexpected reversal of behavior for glass-forming alloys as the temperature of crystallization approaches the glass transition. For this purpose, we simulate the crystallization process in two glass-forming Copper alloys, $Ag_6Cu_4$, which has a positive heat of mixing, and in $CuZr$, characterized by a large negative heat of mixing. Our results allow us to identify that this unusual behavior is directly correlated with a non-monotonic temperature dependence for the formation energy of connected icosahedral structures, which are incompatible with crystalline order and impede the development of the crystal nucleus, leading to an unexpectedly larger free energy barrier at low temperature. This, in turn, promotes the formation of a predominantly closed-packed critical nucleus, with fewer defects, thereby suggesting a new way to control the structure of the crystal nucleus, which is of key importance in catalysis.
\end{abstract}

\maketitle

The formation of crystalline and amorphous solids are key processes in solid-state physics and chemistry~\cite{debenedetti2001supercooled,bernstein2002polymorphism}, with applications in pharmaceuticals, electronics and nanoscience. Crystals have properties that strongly depend on their structure, and can be tailored by modifying the conditions of crystallization~\cite{bernstein2002polymorphism}. Amorphous solids or glasses can be extremely advantageous as their properties can outperform those of their crystalline counterparts, as e.g. amorphous solids of pharmaceutical compounds are often more soluble in water and metallic glasses exhibit improved magnetic and mechanical properties~\cite{hirata2013geometric,schroers2013bulk}. It is therefore crucial to understand how the competition between crystallization and glass formation takes place at the microscopic level to be able to control the type of material obtained.

The microscopic structure of supercooled liquids has become increasingly pictured as inhomogeneous~\cite{tanaka2012bond,hirata2011direct,hirata2013geometric,shen2009icosahedral}. It results from the existence, within the supercooled liquid of short-range order structures, e.g. icosahedral short-range order (ISRO)  as discussed in Frank's pioneering work~\cite{frank1952supercooling}, and of medium-range order (MRO), which are larger domains, either crystal-like~\cite{leocmach2012roles} or composed of connected icosahedra~\cite{sheng2006atomic}. The subtle balance between the two types of order has a complex, yet fascinating, impact on crystallization. On the one hand, icosahedral structures are structurally incompatible with crystalline order and lead to an enhancement of the glass-forming ability~\cite{tanaka2012bond}. On the other hand, amorphous precursors have been shown to promote crystallization during biomineralization~\cite{weiner2005choosing} and fast crystal growth modes have been reported in glasses~\cite{hikima1995determination,konishi2007possible,sun2008crystallization,orava2014fast}. In this work, we focus on the crystal nucleation process, and show that the free energy of nucleation in a metallic glass-forming liquid exhibits a non-monotonic temperature dependence, contrary to the expected behavior~\cite{bernstein2002polymorphism}. To analyze this, we determine the formation energy of connected ISROs~\cite{wu2016critical}, and show that the formation energy of connected ISROs with a large node degree also exhibits a non-monotonic temperature dependence. This correlation between two non-monotonic behaviors sheds new light on the competition between the onset of crystalline order and the formation of connected icosahedra that takes place in a supercooled liquid.

Metallic glasses have recently been used to prepare nanoscale catalytic fibers~\cite{schroers2013bulk}. Similarly, metal nanocrystals have applications in a wide array of fields, e.g. in biology~\cite{langille2012stepwise} and chemistry~\cite{zheng2009observation}. In bimetallic alloys, a competition between glass transition and crystallization can be triggered by the choice of the metals involved. The relative sizes of the two metals plays a major role in the formation of glasses, with very dissimilar radii for the two metals  (with a ratio of 1.1 and above) favoring the onset of a glass transition. Here, we focus on the $Ag_6Cu_4$ and $CuZr$ alloys. Both exhibit a large size ratio ($r_{Ag}:r_{Cu}=1.13$ and $r_{Zr}:r_{Cu}=1.244$) and become a glass below $500$~K for $Ag_6Cu_4$ and $700$~K for $CuZr$~\cite{kimura1998quantum,duan2005molecular}. To examine the interplay between crystallization and glass transition, we study crystal nucleation from increasingly supercooled liquids, i.e. closer and closer to the glass transition. To identify the effect of the glass transition on nucleation, we also examine the impact of supercooling on $CuNi$, for which the size difference is less than 2.5\% and no glass transition is observed. 

We model the interactions between metal atoms using embedded atom model potentials~\cite{kimura1998quantum,duan2005molecular}, that have been shown to model accurately metal properties, with e.g. a predicted melting temperature of $1090$~K for $Ag_6Cu_4$, in excellent agreement with experiments ($1053$~K) and a glass transition around $500$~K~\cite{kimura1998quantum}. We simulate crystal nucleation for increasing degrees of supercooling for $Ag_6Cu_4$, $CuZr$ and $CuNi$. The formation of a crystal nucleus of a critical size is an activated process associated with a large free energy barrier, due to the cost of creating a crystal nucleus within a liquid. To determine this free energy barrier, we combine molecular dynamics simulations with an umbrella sampling ($US$) approach~\cite{tenWolde,PRL}, and the $Q_6$ order parameter that captures the onset of crystalline order~\cite{Gasser,auer2001prediction,desgranges2014unraveling}. This method allows~\cite{auer2001prediction} us to stabilize the system at each step of the nucleation process, enabling the collection of average properties over long simulation runs. The nucleation pathway is typically split into 15 $US$ windows, for which the $US$ potential maintains the system around a fixed value for the order parameter $Q_6$. For each simulation window, we perform a first simulation run to reach the target value for $Q_6$ and to allow the system to relax. Then, once the system has reached a steady-state, we carry out a second production run over which averages are collected~\cite{desgranges2014unraveling}. We perform simulations under isothermal and isobaric conditions ($P=1$~atm). The temperatures of crystallization range from $708.5$~K to $599.5$~K for $Ag_6Cu_4$ (i.e. for degrees of supercooling from 35\% to 45\%), from $840$~K to $728$~K for $CuZr$ (i.e. for degrees of supercooling from 40\% to 48\%) and from $1132.5$~K to $981.5$~K for $CuNi$ (or degrees of supercooling from 25\% to 35\%). The free energy barrier of nucleation, as well as the size and structure of the critical nuclei are determined from the US simulations, as discussed in prior work~\cite{auer2001prediction,desgranges2014unraveling}. The size and structure of the critical nucleus are determined using the criteria defined by Frenkel {\it et al.}~\cite{tenWolde,Auer,PRL,desgranges2014unraveling} during the course of the US simulation window when the system is at the top of the free energy barrier. Specifically, an atom is identified as having a solid-like environment if more than 6 of their nearest neighbors show a highly correlated local environment (i.e. with a dot product of their ${q_{6l}}$ vectors greater than $0.6$)~\cite{desgranges2016effect}. The size of the nucleus is then determined through a cluster analysis. We add that the critical nuclei generated during the $US$ simulations have been checked to be genuine critical nuclei in unbiased simulations~\cite{PRL}.

\begin{figure}
\begin{center}
\includegraphics*[width=8.5cm]{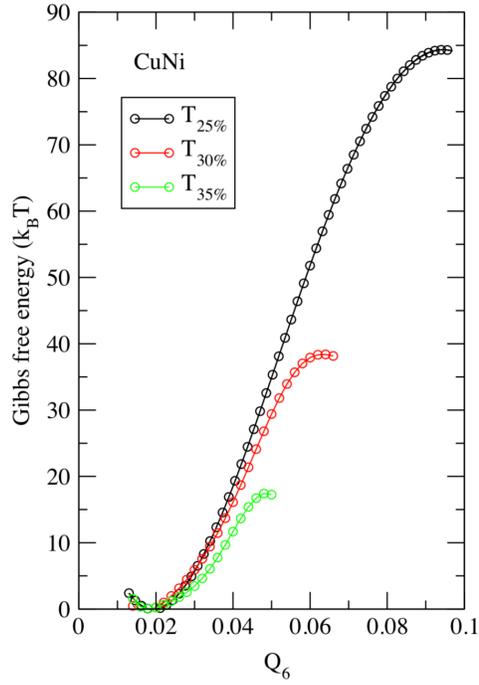}(a)
\includegraphics*[width=10cm]{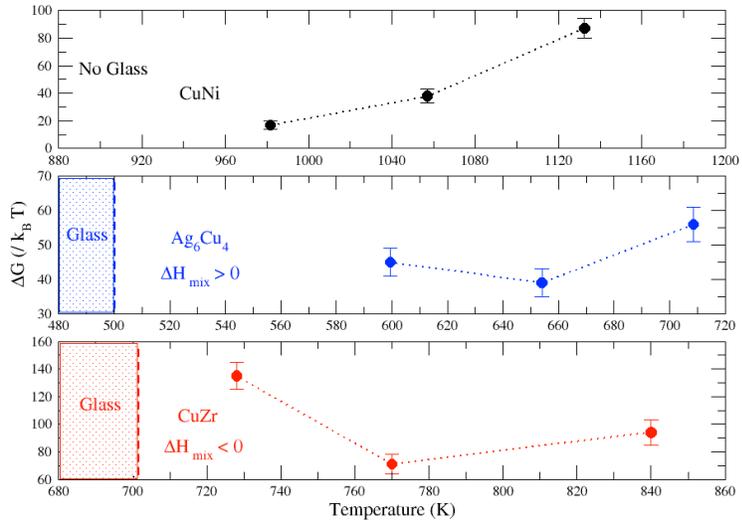}(b)
\end{center}
\caption{(a) Gibbs free energy profiles of nucleation for liquid $CuNi$ for different degrees of supercooling, and (b) $\Delta G$ as a function of temperature for $CuNi$ and for the two glass-forming Copper alloys $Ag_6Cu_4$ and $CuZr$.}
\label{Fig1}
\end{figure}

We first discuss the results obtained for the free energy of nucleation for $CuNi$. Fig.~\ref{Fig1}(a) shows the free energy profile of nucleation for degrees of supercooling ranging from 25\% to 35\%. For each temperature, we gradually increase the imposed value for $Q_6$ and observe, as a result, the formation of a crystal nucleus. When the system reaches the top of the free energy barrier, we obtain a critical nucleus that can either continue to grow, or dissolve back into the liquid. As shown in Fig.~\ref{Fig1}(a), increasing the degree of supercooling results in the following trend. We observe that the height of the free energy barrier is higher for a low supercooling, with a free energy of nucleation estimated at $87\pm7~k_BT$ for $T_{25\%}$, at $38\pm5~k_BT$ for $T_{30\%}$, and at $17\pm3~k_BT$ for $T_{35\%}$. Furthermore, the monotonic variation, found for the nucleation free energy with the degree of supercooling, is in line with the predictions from the classical nucleation theory~\cite{bernstein2002polymorphism} and with prior simulation results on binary ionic systems~\cite{valeriani2005rate} and pure metals~\cite{desgranges2007molecular}. Turning to the results obtained for the two glass-forming copper alloys, we observe a dramatically different behavior in Fig.~\ref{Fig1}(b). Unlike for $CuNi$, we find that the free energy of nucleation exhibit non-monotonic variations with the degree of supercooling. For $Ag_6Cu_4$, the free energy of nucleation starts by decreasing from $56\pm5~k_BT$ ($T_{35\%}$) to $39\pm4~k_BT$  ($T_{40\%}$), before increasing again to $45\pm4~k_BT$ for $T_{45\%}$. Similarly, Fig.~\ref{Fig1}(b) shows that, for the $CuZr$ alloy, the free energy of nucleation also exhibits a non-monotonic dependence on temperature, as it first decreases from $94\pm9~k_BT$ ($T_{40\%}$) to $71\pm7~k_BT$  ($T_{45\%}$), before increasing again to $135\pm10~k_BT$ for $T_{48\%}$. This behavior strongly departs from the predictions of the classical nucleation theory, and from the results for the non-glass forming $CuNi$. This shows that, as the conditions of crystallization for $Ag_6Cu_4$ and $CuZr$ get closer to the glass transition, we observe an unusual, non-monotonic, crystallization behavior.

\begin{figure}
\begin{center}
\includegraphics*[width=14cm]{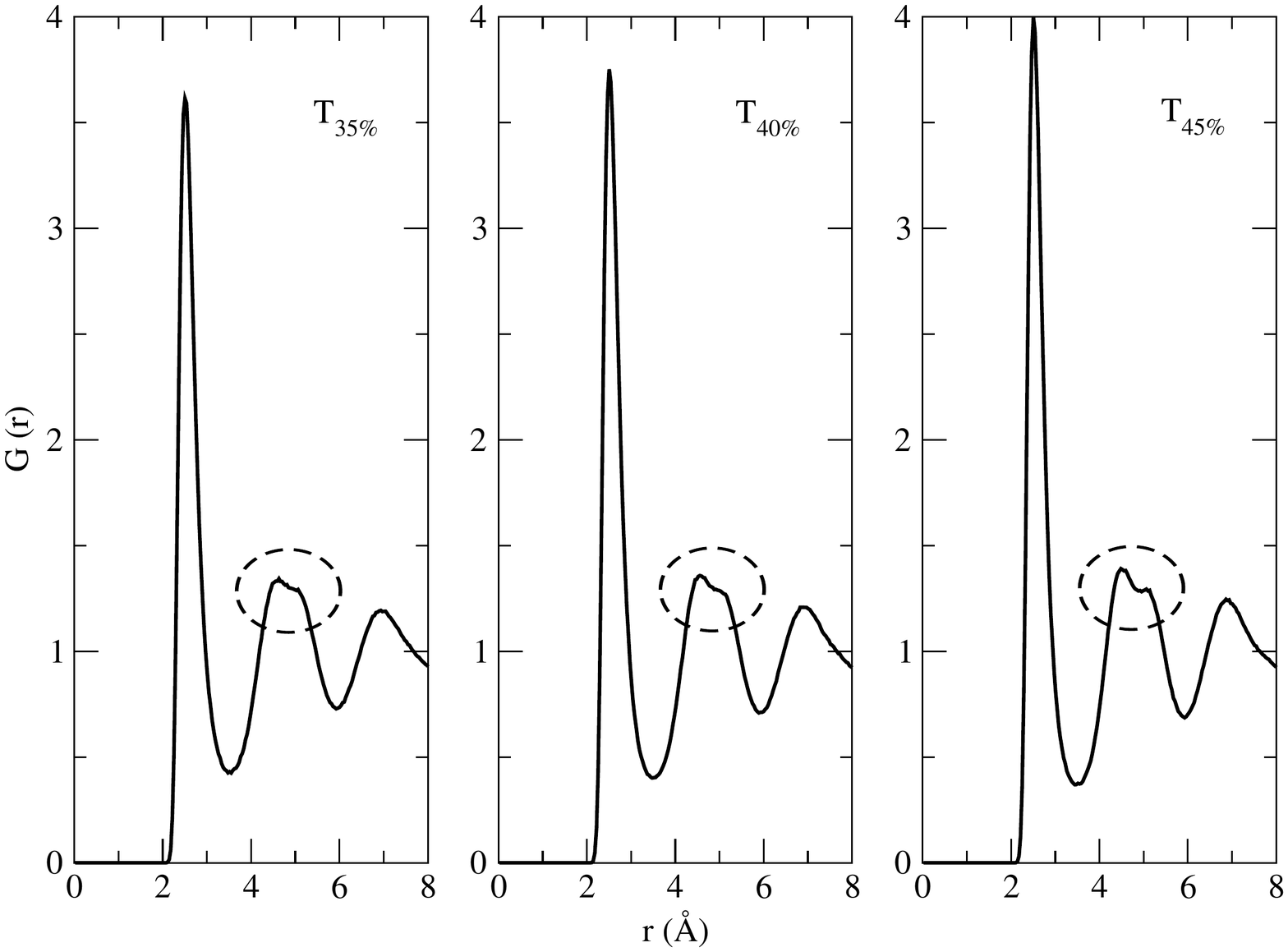}
\end{center}
\caption{Copper-Copper pair distribution function $G(r)$ in supercooled $Ag_6Cu_4$ liquids for a supercooling of 35\% (left), 40\% (middle) and 45\% (right). The circled regions show the splitting of the second maximum in $G(r)$.}
\label{Fig2}
\end{figure}

\begin{figure}
\begin{center}
\includegraphics*[width=6.8cm]{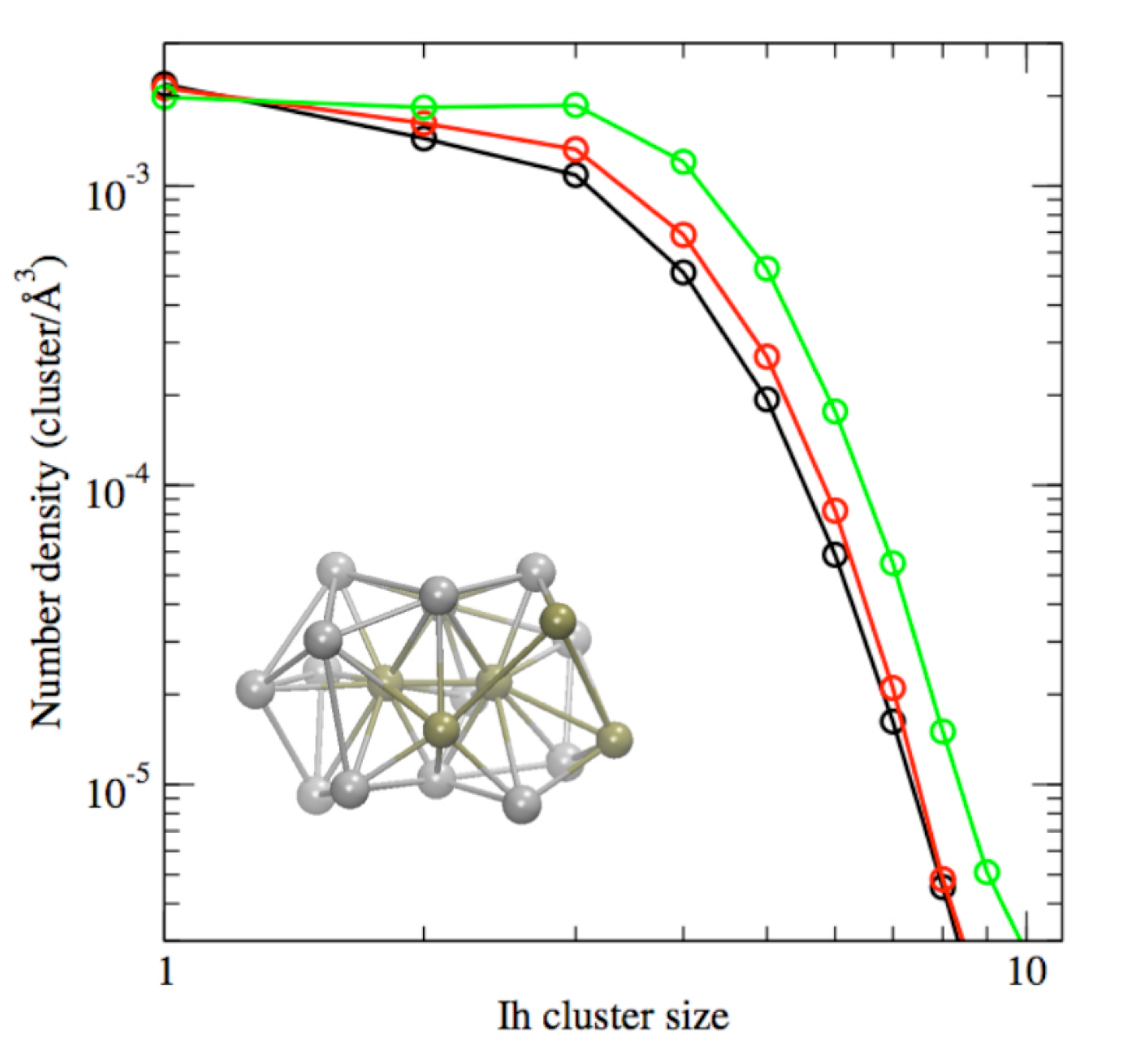}(a)
\includegraphics*[width=8.2cm]{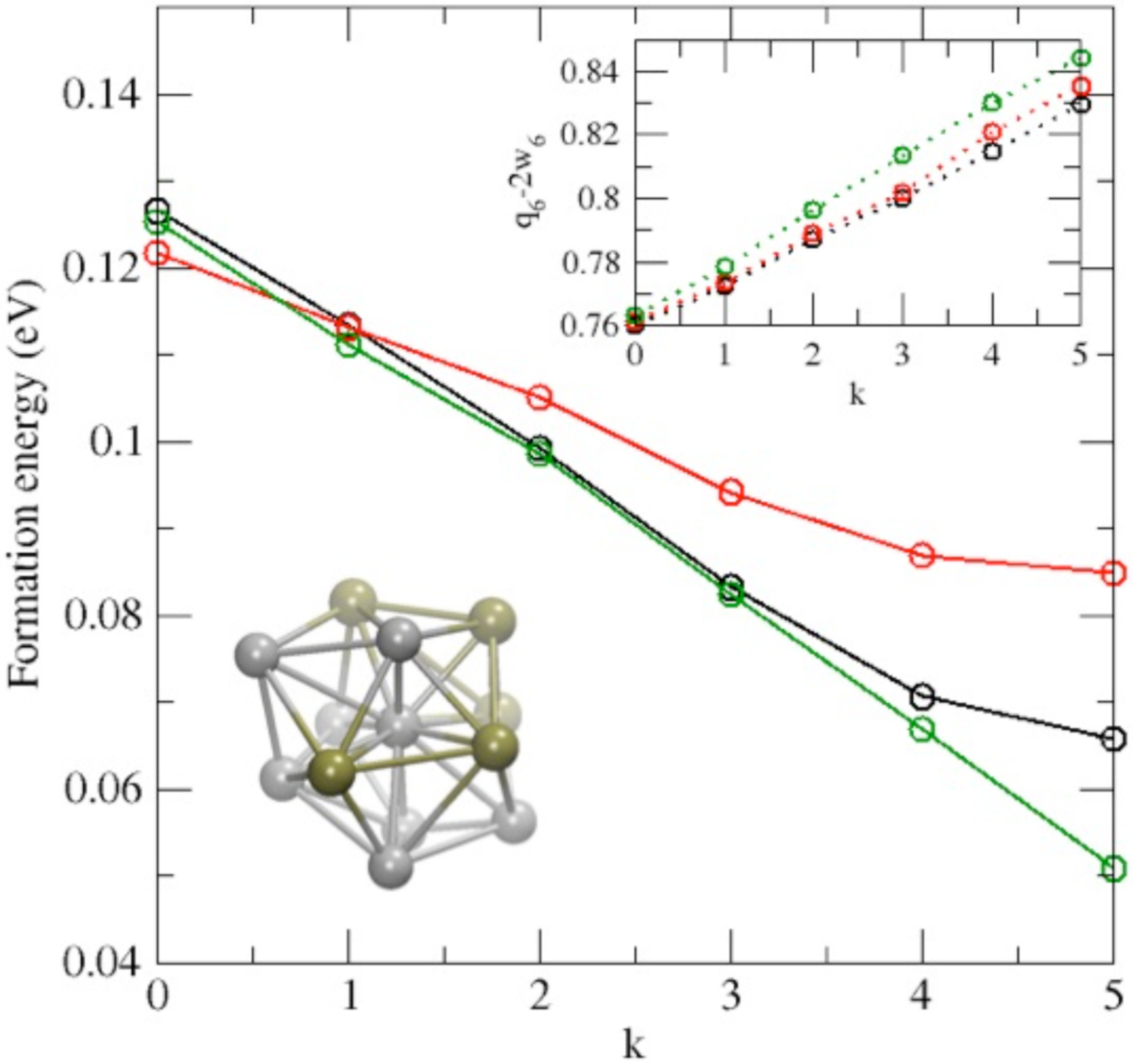}(b)
\end{center}
\caption{Supercooled $Ag_6Cu_4$. (a) Number density distributions of clusters formed by connected ISROs via volume sharing at $T_{35\%}$ (black), $T_{40\%}$ (red) and $T_{45\%}$ (green), with a snapshot showing 2 connected ISROs. (b) ISRO formation energy as a function of the node degree $k$, with a snapshot showing a configuration for $k=0$. The inset shows the increase in the Ih order parameter with $k$.}
\label{Fig3}
\end{figure}

To analyze this, we examine the effect of supercooling on the parent phase and show in Fig.~\ref{Fig2} the pair distribution functions ($G(r)$) in the metastable liquid for $Ag_6Cu_4$ in the absence of any $US$ potential. $G(r)$ displays, as $T$ decreases, the expected increase in the height of the first peak, characteristic of the increase in SRO, coupled with a decrease in the minimum that immediately follows. These functions also reveal a subtle change in $G(r)$, with a splitting of the second peak at low temperature. The same behavior has also been reported~\cite{duan2005molecular} in the case of $CuZr$. This indicates that the system approaches the glass transition~\cite{kimura1998quantum}, and points to the development of MRO in the liquid~\cite{liang2014influence}. This stems from the increased geometric frustration~\cite{tanaka2012bond}, as non space-filling clusters, such as e.g. icosahedral short-range order (ISRO), form in the supercooled liquid and prevent the onset of crystal nucleation. Recent experiments on metallic glasses~\cite{shen2009icosahedral} have provided support for an icosahedral order-based frustration model in such systems. Furthermore, the increase in icosahedral order has been linked to the splitting of the second peak of $G(r)$~\cite{liang2014influence}. We therefore start by quantifying the ISROs~\cite{desgranges2007molecular}. Following the analysis of Wu {\it et al.}~\cite{wu2016critical}, we find that the presence of ISROs goes on to impact the MRO in the supercooled liquid, with the increase, as temperature decreases, of connections between ISROs via volume sharing (see Fig.~\ref{Fig3}(a)). Upon closer inspection, Fig.~\ref{Fig3}(a) shows that a decrease from $T_{40\%}$ to $T_{45\%}$ results in a much more significant increase in the number density of larger clusters of connected ISROs, than the decrease from $T_{35\%}$ to $T_{40\%}$. To understand this, we calculate in Fig.~\ref{Fig3}(b) the ISRO formation energy as a function of the node degree $k$, i.e. the number of ISROs directly connected to it. In this calculation, we set the reference for the energy of each atom to its value in the $FCC$ crystal for $Ag$ and $Cu$, as in previous work~\cite{wu2016critical}. This plot shows a decrease in the formation energy as $k$ increases, in line with prior work on supercooled $CuZr$. Turning to the temperature dependence of the results, we find that the formation energy for larger $k$ values increases when $T$ decreases from $T_{35\%}$ to $T_{40\%}$. Since fewer connected ISROs are expected to favor crystallization, this greater formation energy for connected ISROs at $T_{40\%}$ is consistent with the decrease in the free energy of nucleation found in Fig.~\ref{Fig1}(b). On the other hand, the formation energy for the larger $k$ undergoes a significant decrease with $T$ going from $T_{40\%}$ to $T_{45\%}$. This means that multiple connections between ISROs form much more easily at $T_{45\%}$. Since connected ISROs lead to a MRO incompatible with crystalline order, the much lower formation energy for $T_{45\%}$ is also consistent with the increase in the free energy of nucleation at this temperature. These results therefore establish a direct correlation between the non-monotonic temperature dependence of the free energy of nucleation and the non-monotonic temperature dependence of the formation energy of connected ISROs for $k>2$.

\begin{figure}
\begin{center}
\includegraphics*[width=12cm]{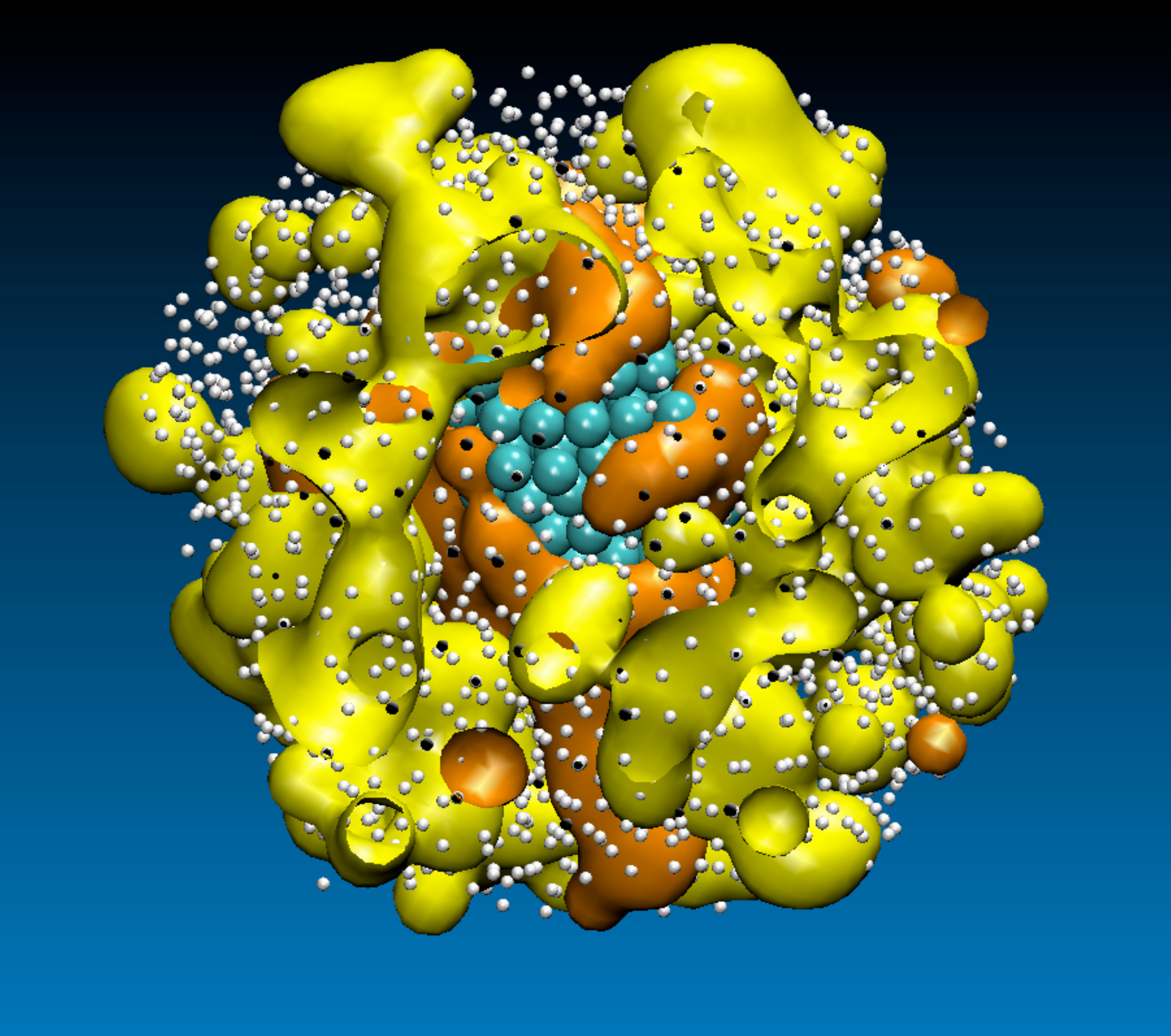}
\end{center}
\caption{Formation of the critical nucleus for the $Ag_6Cu_4$ system ($T_{45\%}$). The crystal nucleus (cyan) is surrounded by a crystal-like ordered region (orange), by ISROs (yellow) and liquid-like atoms (white).}
\label{Fig4}
\end{figure}

How does this impact the nucleation process? As shown in Fig.~\ref{Fig4}, the crystal nucleus is surrounded by a supercooled liquid that exhibits two different types of order. These correspond to the connected ISROs and to a crystal-like region, characterized by large Steinhardt order parameters~\cite{leocmach2012roles}, around the crystal nucleus~\cite{tanaka2012bond}. The presence of connected ISROs hinders the nucleation process, and results in a larger free energy barrier for the nucleation process. We add that the density of ISROs around the crystal nucleus is the same as in the bulk, and that no enrichment in ISROs occurs during crystal nucleation. Finally, we find much lower number densities of ISROs for $CuNi$, which ultimately result in the usual monotonic temperature dependence for the nucleation free energy in the non glass-former $CuNi$.

\begin{figure}
\begin{center}
\includegraphics*[width=10cm]{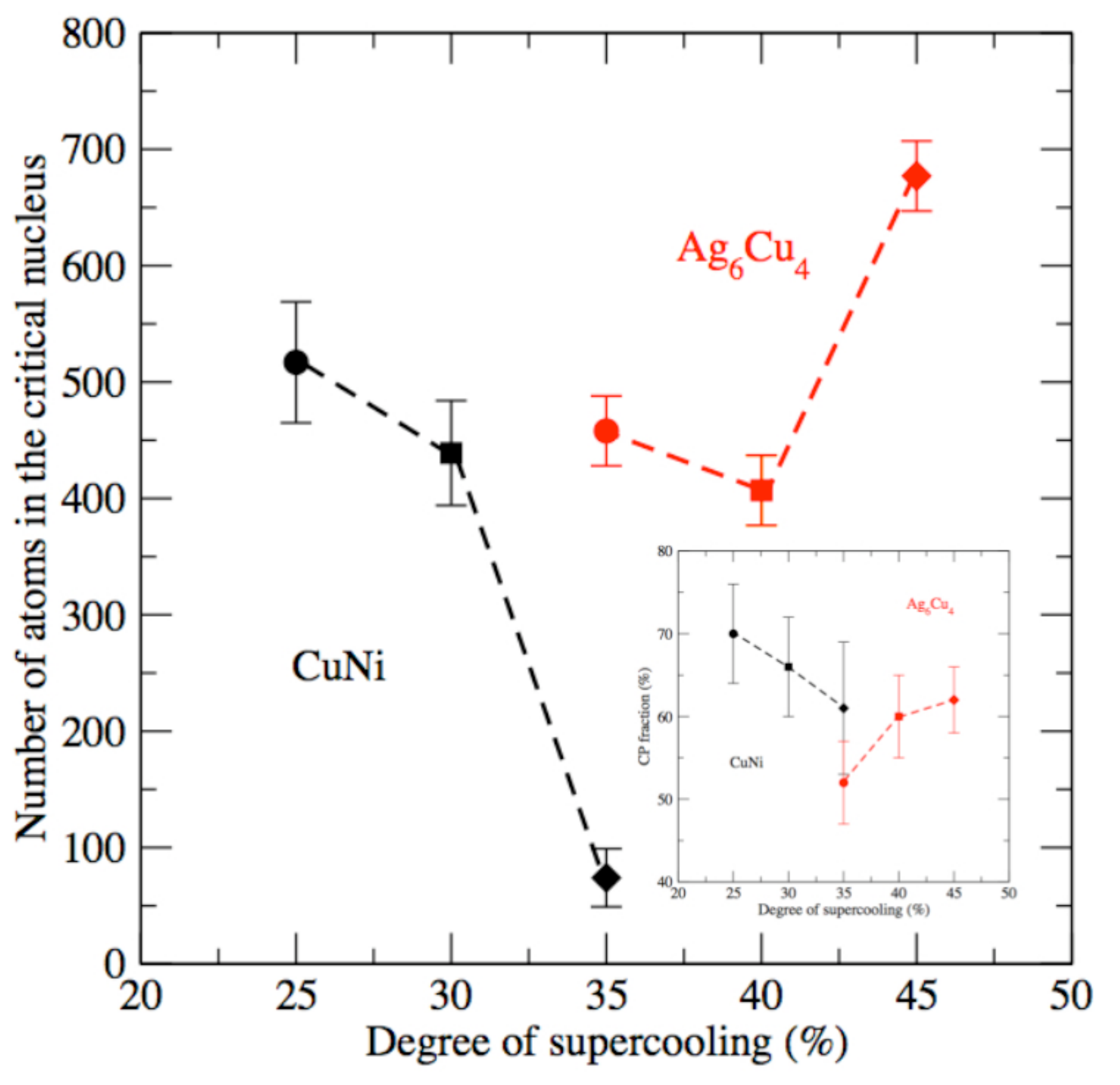}
\end{center}
\caption{Variation of the size of the critical nucleus as a function of the degree of supercooling. (Inset) Structure of the critical nucleus against the degree of supercooling.}
\label{Fig5}
\end{figure}

We now turn to the size and structure of the critical nucleus in Fig.~\ref{Fig5} and focus for comparison puporses on the two alloys, $CuNi$ and $Ag_6Cu_4$, for which both elements form FCC crystals. For $CuNi$, the critical size decreases steadily from $517 \pm 40$ atoms ($T_{25\%}$) to $74 \pm 15$ atoms ($T_{35\%}$). This monotonic behavior is in line with the predictions from the classical nucleation theory. Fig.~\ref{Fig5} also shows that the fraction in the stable close-packed ($CP$) structure decreases as the degree of supercooling increases, since all polymorphs form easily at low temperature. For $Ag_6Cu_4$, we observe very different behaviors. First, the critical size decreases from $458 \pm 30$~atoms ($T_{35\%}$) to $407 \pm 25$~atoms ($T_{40\%}$) and then increases back to $677 \pm 40$ ($T_{45\%}$). This non-monotonic behavior is consistent with the variations observed for the nucleation free energy. Second, the structure becomes more dominated by the stable $CP$ structure as the degree of supercooling increases. This is likely a consequence of the role played by the ISROs surrounding the crystal nucleus, which slow down nucleation and lead to a nucleus with fewer defects. This increase in $CP$ fraction correlates well with the increase in the density of ISROs in the supercooled liquid, thereby providing support for this mechanism. It also suggests a new way of controlling polymorphism by setting the conditions of crystallization close to the glass transition.

Our results show that an unexpected crystallization behavior arises, as the system approaches the glass transition. Close to this transition, a further decrease in temperature results in a reversal of crystallization behavior, with the formation of a larger critical nucleus associated with a larger free energy barrier of nucleation. This unusual behavior is directly correlated with a non-monotonic temperature dependence for the formation energy of connected icosahedral structures. Given the existence of a glass transition for molecular and pharmaceutical compounds~\cite{sun2008crystallization}, the conclusions drawn here extend to a wide range of systems, well beyond the case of metallic systems.\\

{\bf Acknowledgements}
Partial funding for this research was provided by the National Science Foundation (NSF) through CAREER award DMR-1052808.\\

\bibliography{Glacryst}

\begin{thebibliography}{30}
\expandafter\ifx\csname natexlab\endcsname\relax\def\natexlab#1{#1}\fi
\expandafter\ifx\csname bibnamefont\endcsname\relax
  \def\bibnamefont#1{#1}\fi
\expandafter\ifx\csname bibfnamefont\endcsname\relax
  \def\bibfnamefont#1{#1}\fi
\expandafter\ifx\csname citenamefont\endcsname\relax
  \def\citenamefont#1{#1}\fi
\expandafter\ifx\csname url\endcsname\relax
  \def\url#1{\texttt{#1}}\fi
\expandafter\ifx\csname urlprefix\endcsname\relax\def\urlprefix{URL }\fi
\providecommand{\bibinfo}[2]{#2}
\providecommand{\eprint}[2][]{\url{#2}}

\bibitem[{\citenamefont{Debenedetti and
  Stillinger}(2001)}]{debenedetti2001supercooled}
\bibinfo{author}{\bibfnamefont{P.~G.} \bibnamefont{Debenedetti}}
  \bibnamefont{and} \bibinfo{author}{\bibfnamefont{F.~H.}
  \bibnamefont{Stillinger}}, \bibinfo{journal}{Nature}
  \textbf{\bibinfo{volume}{410}}, \bibinfo{pages}{259} (\bibinfo{year}{2001}).

\bibitem[{\citenamefont{Bernstein}(2002)}]{bernstein2002polymorphism}
\bibinfo{author}{\bibfnamefont{J.}~\bibnamefont{Bernstein}},
  \emph{\bibinfo{title}{Polymorphism in molecular crystals}}
  (\bibinfo{publisher}{Clarendon Press, Oxford}, \bibinfo{year}{2002}).

\bibitem[{\citenamefont{Hirata et~al.}(2013)\citenamefont{Hirata, Kang, Fujita,
  Klumov, Matsue, Kotani, Yavari, and Chen}}]{hirata2013geometric}
\bibinfo{author}{\bibfnamefont{A.}~\bibnamefont{Hirata}},
  \bibinfo{author}{\bibfnamefont{L.}~\bibnamefont{Kang}},
  \bibinfo{author}{\bibfnamefont{T.}~\bibnamefont{Fujita}},
  \bibinfo{author}{\bibfnamefont{B.}~\bibnamefont{Klumov}},
  \bibinfo{author}{\bibfnamefont{K.}~\bibnamefont{Matsue}},
  \bibinfo{author}{\bibfnamefont{M.}~\bibnamefont{Kotani}},
  \bibinfo{author}{\bibfnamefont{A.}~\bibnamefont{Yavari}}, \bibnamefont{and}
  \bibinfo{author}{\bibfnamefont{M.}~\bibnamefont{Chen}},
  \bibinfo{journal}{Science} \textbf{\bibinfo{volume}{341}},
  \bibinfo{pages}{376} (\bibinfo{year}{2013}).

\bibitem[{\citenamefont{Schroers}(2013)}]{schroers2013bulk}
\bibinfo{author}{\bibfnamefont{J.}~\bibnamefont{Schroers}},
  \bibinfo{journal}{Physics Today} \textbf{\bibinfo{volume}{66}},
  \bibinfo{pages}{32} (\bibinfo{year}{2013}).

\bibitem[{\citenamefont{Tanaka}(2012)}]{tanaka2012bond}
\bibinfo{author}{\bibfnamefont{H.}~\bibnamefont{Tanaka}},
  \bibinfo{journal}{Eur. Phys. J. E} \textbf{\bibinfo{volume}{35}},
  \bibinfo{pages}{113} (\bibinfo{year}{2012}).

\bibitem[{\citenamefont{Hirata et~al.}(2011)\citenamefont{Hirata, Guan, Fujita,
  Hirotsu, Inoue, Yavari, Sakurai, and Chen}}]{hirata2011direct}
\bibinfo{author}{\bibfnamefont{A.}~\bibnamefont{Hirata}},
  \bibinfo{author}{\bibfnamefont{P.}~\bibnamefont{Guan}},
  \bibinfo{author}{\bibfnamefont{T.}~\bibnamefont{Fujita}},
  \bibinfo{author}{\bibfnamefont{Y.}~\bibnamefont{Hirotsu}},
  \bibinfo{author}{\bibfnamefont{A.}~\bibnamefont{Inoue}},
  \bibinfo{author}{\bibfnamefont{A.~R.} \bibnamefont{Yavari}},
  \bibinfo{author}{\bibfnamefont{T.}~\bibnamefont{Sakurai}}, \bibnamefont{and}
  \bibinfo{author}{\bibfnamefont{M.}~\bibnamefont{Chen}},
  \bibinfo{journal}{Nat. Mater.} \textbf{\bibinfo{volume}{10}},
  \bibinfo{pages}{28} (\bibinfo{year}{2011}).

\bibitem[{\citenamefont{Shen et~al.}(2009)\citenamefont{Shen, Kim,
  Gangopadhyay, and Kelton}}]{shen2009icosahedral}
\bibinfo{author}{\bibfnamefont{Y.}~\bibnamefont{Shen}},
  \bibinfo{author}{\bibfnamefont{T.}~\bibnamefont{Kim}},
  \bibinfo{author}{\bibfnamefont{A.}~\bibnamefont{Gangopadhyay}},
  \bibnamefont{and} \bibinfo{author}{\bibfnamefont{K.}~\bibnamefont{Kelton}},
  \bibinfo{journal}{Phys. Rev. Lett.} \textbf{\bibinfo{volume}{102}},
  \bibinfo{pages}{057801} (\bibinfo{year}{2009}).

\bibitem[{\citenamefont{Frank}(1952)}]{frank1952supercooling}
\bibinfo{author}{\bibfnamefont{F.}~\bibnamefont{Frank}},
  \bibinfo{journal}{Proc. R. Soc. A} pp. \bibinfo{pages}{43--46}
  (\bibinfo{year}{1952}).

\bibitem[{\citenamefont{Leocmach and Tanaka}(2012)}]{leocmach2012roles}
\bibinfo{author}{\bibfnamefont{M.}~\bibnamefont{Leocmach}} \bibnamefont{and}
  \bibinfo{author}{\bibfnamefont{H.}~\bibnamefont{Tanaka}},
  \bibinfo{journal}{Nat. Commun.} \textbf{\bibinfo{volume}{3}},
  \bibinfo{pages}{974} (\bibinfo{year}{2012}).

\bibitem[{\citenamefont{Sheng et~al.}(2006)\citenamefont{Sheng, Luo, Alamgir,
  Bai, and Ma}}]{sheng2006atomic}
\bibinfo{author}{\bibfnamefont{H.}~\bibnamefont{Sheng}},
  \bibinfo{author}{\bibfnamefont{W.}~\bibnamefont{Luo}},
  \bibinfo{author}{\bibfnamefont{F.}~\bibnamefont{Alamgir}},
  \bibinfo{author}{\bibfnamefont{J.}~\bibnamefont{Bai}}, \bibnamefont{and}
  \bibinfo{author}{\bibfnamefont{E.}~\bibnamefont{Ma}},
  \bibinfo{journal}{Nature} \textbf{\bibinfo{volume}{439}},
  \bibinfo{pages}{419} (\bibinfo{year}{2006}).

\bibitem[{\citenamefont{Weiner et~al.}(2005)\citenamefont{Weiner, Sagi, and
  Addadi}}]{weiner2005choosing}
\bibinfo{author}{\bibfnamefont{S.}~\bibnamefont{Weiner}},
  \bibinfo{author}{\bibfnamefont{I.}~\bibnamefont{Sagi}}, \bibnamefont{and}
  \bibinfo{author}{\bibfnamefont{L.}~\bibnamefont{Addadi}},
  \bibinfo{journal}{Science} \textbf{\bibinfo{volume}{309}},
  \bibinfo{pages}{1027} (\bibinfo{year}{2005}).

\bibitem[{\citenamefont{Hikima et~al.}(1995)\citenamefont{Hikima, Adachi,
  Hanaya, and Oguni}}]{hikima1995determination}
\bibinfo{author}{\bibfnamefont{T.}~\bibnamefont{Hikima}},
  \bibinfo{author}{\bibfnamefont{Y.}~\bibnamefont{Adachi}},
  \bibinfo{author}{\bibfnamefont{M.}~\bibnamefont{Hanaya}}, \bibnamefont{and}
  \bibinfo{author}{\bibfnamefont{M.}~\bibnamefont{Oguni}},
  \bibinfo{journal}{Phys. Rev. B} \textbf{\bibinfo{volume}{52}},
  \bibinfo{pages}{3900} (\bibinfo{year}{1995}).

\bibitem[{\citenamefont{Konishi and Tanaka}(2007)}]{konishi2007possible}
\bibinfo{author}{\bibfnamefont{T.}~\bibnamefont{Konishi}} \bibnamefont{and}
  \bibinfo{author}{\bibfnamefont{H.}~\bibnamefont{Tanaka}},
  \bibinfo{journal}{Phys. Rev. B} \textbf{\bibinfo{volume}{76}},
  \bibinfo{pages}{220201} (\bibinfo{year}{2007}).

\bibitem[{\citenamefont{Sun et~al.}(2008)\citenamefont{Sun, Xi, Chen, Ediger,
  and Yu}}]{sun2008crystallization}
\bibinfo{author}{\bibfnamefont{Y.}~\bibnamefont{Sun}},
  \bibinfo{author}{\bibfnamefont{H.}~\bibnamefont{Xi}},
  \bibinfo{author}{\bibfnamefont{S.}~\bibnamefont{Chen}},
  \bibinfo{author}{\bibfnamefont{M.}~\bibnamefont{Ediger}}, \bibnamefont{and}
  \bibinfo{author}{\bibfnamefont{L.}~\bibnamefont{Yu}}, \bibinfo{journal}{J.
  Phys. Chem. B} \textbf{\bibinfo{volume}{112}}, \bibinfo{pages}{5594}
  (\bibinfo{year}{2008}).

\bibitem[{\citenamefont{Orava and Greer}(2014)}]{orava2014fast}
\bibinfo{author}{\bibfnamefont{J.}~\bibnamefont{Orava}} \bibnamefont{and}
  \bibinfo{author}{\bibfnamefont{A.~L.} \bibnamefont{Greer}},
  \bibinfo{journal}{J. Chem. Phys.} \textbf{\bibinfo{volume}{140}},
  \bibinfo{pages}{214504} (\bibinfo{year}{2014}).

\bibitem[{\citenamefont{Wu et~al.}(2016)\citenamefont{Wu, Li, Huo, Li, Wang,
  and Liu}}]{wu2016critical}
\bibinfo{author}{\bibfnamefont{Z.}~\bibnamefont{Wu}},
  \bibinfo{author}{\bibfnamefont{F.}~\bibnamefont{Li}},
  \bibinfo{author}{\bibfnamefont{C.}~\bibnamefont{Huo}},
  \bibinfo{author}{\bibfnamefont{M.}~\bibnamefont{Li}},
  \bibinfo{author}{\bibfnamefont{W.}~\bibnamefont{Wang}}, \bibnamefont{and}
  \bibinfo{author}{\bibfnamefont{K.}~\bibnamefont{Liu}}, \bibinfo{journal}{Sci.
  Rep.} \textbf{\bibinfo{volume}{6}} (\bibinfo{year}{2016}).

\bibitem[{\citenamefont{Langille et~al.}(2012)\citenamefont{Langille, Zhang,
  Personick, Li, and Mirkin}}]{langille2012stepwise}
\bibinfo{author}{\bibfnamefont{M.~R.} \bibnamefont{Langille}},
  \bibinfo{author}{\bibfnamefont{J.}~\bibnamefont{Zhang}},
  \bibinfo{author}{\bibfnamefont{M.~L.} \bibnamefont{Personick}},
  \bibinfo{author}{\bibfnamefont{S.}~\bibnamefont{Li}}, \bibnamefont{and}
  \bibinfo{author}{\bibfnamefont{C.~A.} \bibnamefont{Mirkin}},
  \bibinfo{journal}{Science} \textbf{\bibinfo{volume}{337}},
  \bibinfo{pages}{954} (\bibinfo{year}{2012}).

\bibitem[{\citenamefont{Zheng et~al.}(2009)\citenamefont{Zheng, Smith, Jun,
  Kisielowski, Dahmen, and Alivisatos}}]{zheng2009observation}
\bibinfo{author}{\bibfnamefont{H.}~\bibnamefont{Zheng}},
  \bibinfo{author}{\bibfnamefont{R.~K.} \bibnamefont{Smith}},
  \bibinfo{author}{\bibfnamefont{Y.-w.} \bibnamefont{Jun}},
  \bibinfo{author}{\bibfnamefont{C.}~\bibnamefont{Kisielowski}},
  \bibinfo{author}{\bibfnamefont{U.}~\bibnamefont{Dahmen}}, \bibnamefont{and}
  \bibinfo{author}{\bibfnamefont{A.~P.} \bibnamefont{Alivisatos}},
  \bibinfo{journal}{Science} \textbf{\bibinfo{volume}{324}},
  \bibinfo{pages}{1309} (\bibinfo{year}{2009}).

\bibitem[{\citenamefont{Kimura et~al.}(1999)\citenamefont{Kimura, Qi, Cagin,
  and Goddard}}]{kimura1998quantum}
\bibinfo{author}{\bibfnamefont{Y.}~\bibnamefont{Kimura}},
  \bibinfo{author}{\bibfnamefont{Y.}~\bibnamefont{Qi}},
  \bibinfo{author}{\bibfnamefont{T.}~\bibnamefont{Cagin}}, \bibnamefont{and}
  \bibinfo{author}{\bibfnamefont{W.~A.} \bibnamefont{Goddard}},
  \bibinfo{journal}{Phys. Rev. B} \textbf{\bibinfo{volume}{59}},
  \bibinfo{pages}{3527} (\bibinfo{year}{1999}).

\bibitem[{\citenamefont{Duan et~al.}(2005)\citenamefont{Duan, Xu, Zhang, Zhang,
  Cagin, Johnson, and Goddard~III}}]{duan2005molecular}
\bibinfo{author}{\bibfnamefont{G.}~\bibnamefont{Duan}},
  \bibinfo{author}{\bibfnamefont{D.}~\bibnamefont{Xu}},
  \bibinfo{author}{\bibfnamefont{Q.}~\bibnamefont{Zhang}},
  \bibinfo{author}{\bibfnamefont{G.}~\bibnamefont{Zhang}},
  \bibinfo{author}{\bibfnamefont{T.}~\bibnamefont{Cagin}},
  \bibinfo{author}{\bibfnamefont{W.~L.} \bibnamefont{Johnson}},
  \bibnamefont{and} \bibinfo{author}{\bibfnamefont{W.~A.}
  \bibnamefont{Goddard~III}}, \bibinfo{journal}{Phys. Rev. B}
  \textbf{\bibinfo{volume}{71}}, \bibinfo{pages}{224208}
  (\bibinfo{year}{2005}).

\bibitem[{\citenamefont{ten Wolde et~al.}(1995)\citenamefont{ten Wolde,
  Ruiz-Montero, and Frenkel}}]{tenWolde}
\bibinfo{author}{\bibfnamefont{P.~R.} \bibnamefont{ten Wolde}},
  \bibinfo{author}{\bibfnamefont{M.~J.} \bibnamefont{Ruiz-Montero}},
  \bibnamefont{and} \bibinfo{author}{\bibfnamefont{D.}~\bibnamefont{Frenkel}},
  \bibinfo{journal}{Phys.\ Rev.\ Lett.} \textbf{\bibinfo{volume}{75}},
  \bibinfo{pages}{2714} (\bibinfo{year}{1995}).

\bibitem[{\citenamefont{Desgranges and Delhommelle}(2007{\natexlab{a}})}]{PRL}
\bibinfo{author}{\bibfnamefont{C.}~\bibnamefont{Desgranges}} \bibnamefont{and}
  \bibinfo{author}{\bibfnamefont{J.}~\bibnamefont{Delhommelle}},
  \bibinfo{journal}{Phys. Rev. Lett.} \textbf{\bibinfo{volume}{98}},
  \bibinfo{pages}{235502} (\bibinfo{year}{2007}{\natexlab{a}}).

\bibitem[{\citenamefont{Gasser et~al.}(2001)\citenamefont{Gasser, Weeks,
  Schofield, Pusey, and Weitz}}]{Gasser}
\bibinfo{author}{\bibfnamefont{U.}~\bibnamefont{Gasser}},
  \bibinfo{author}{\bibfnamefont{E.~R.} \bibnamefont{Weeks}},
  \bibinfo{author}{\bibfnamefont{A.}~\bibnamefont{Schofield}},
  \bibinfo{author}{\bibfnamefont{P.~N.} \bibnamefont{Pusey}}, \bibnamefont{and}
  \bibinfo{author}{\bibfnamefont{D.~A.} \bibnamefont{Weitz}},
  \bibinfo{journal}{Science} \textbf{\bibinfo{volume}{292}},
  \bibinfo{pages}{258} (\bibinfo{year}{2001}).

\bibitem[{\citenamefont{Auer and
  Frenkel}(2001{\natexlab{a}})}]{auer2001prediction}
\bibinfo{author}{\bibfnamefont{S.}~\bibnamefont{Auer}} \bibnamefont{and}
  \bibinfo{author}{\bibfnamefont{D.}~\bibnamefont{Frenkel}},
  \bibinfo{journal}{Nature} \textbf{\bibinfo{volume}{409}},
  \bibinfo{pages}{1020} (\bibinfo{year}{2001}{\natexlab{a}}).

\bibitem[{\citenamefont{Desgranges and
  Delhommelle}(2014)}]{desgranges2014unraveling}
\bibinfo{author}{\bibfnamefont{C.}~\bibnamefont{Desgranges}} \bibnamefont{and}
  \bibinfo{author}{\bibfnamefont{J.}~\bibnamefont{Delhommelle}},
  \bibinfo{journal}{J. Am. Chem. Soc.} \textbf{\bibinfo{volume}{136}},
  \bibinfo{pages}{8145} (\bibinfo{year}{2014}).

\bibitem[{\citenamefont{Auer and Frenkel}(2001{\natexlab{b}})}]{Auer}
\bibinfo{author}{\bibfnamefont{S.}~\bibnamefont{Auer}} \bibnamefont{and}
  \bibinfo{author}{\bibfnamefont{D.}~\bibnamefont{Frenkel}},
  \bibinfo{journal}{Nature} \textbf{\bibinfo{volume}{409}},
  \bibinfo{pages}{1023} (\bibinfo{year}{2001}{\natexlab{b}}).

\bibitem[{\citenamefont{Desgranges and
  Delhommelle}(2016)}]{desgranges2016effect}
\bibinfo{author}{\bibfnamefont{C.}~\bibnamefont{Desgranges}} \bibnamefont{and}
  \bibinfo{author}{\bibfnamefont{J.}~\bibnamefont{Delhommelle}},
  \bibinfo{journal}{J. Phys. Chem. C} \textbf{\bibinfo{volume}{120}},
  \bibinfo{pages}{27657} (\bibinfo{year}{2016}).

\bibitem[{\citenamefont{Valeriani et~al.}(2005)\citenamefont{Valeriani, Sanz,
  and Frenkel}}]{valeriani2005rate}
\bibinfo{author}{\bibfnamefont{C.}~\bibnamefont{Valeriani}},
  \bibinfo{author}{\bibfnamefont{E.}~\bibnamefont{Sanz}}, \bibnamefont{and}
  \bibinfo{author}{\bibfnamefont{D.}~\bibnamefont{Frenkel}},
  \bibinfo{journal}{J. Chem. Phys.} \textbf{\bibinfo{volume}{122}},
  \bibinfo{pages}{194501} (\bibinfo{year}{2005}).

\bibitem[{\citenamefont{Desgranges and
  Delhommelle}(2007{\natexlab{b}})}]{desgranges2007molecular}
\bibinfo{author}{\bibfnamefont{C.}~\bibnamefont{Desgranges}} \bibnamefont{and}
  \bibinfo{author}{\bibfnamefont{J.}~\bibnamefont{Delhommelle}},
  \bibinfo{journal}{J. Chem. Phys.} \textbf{\bibinfo{volume}{127}},
  \bibinfo{pages}{144509} (\bibinfo{year}{2007}{\natexlab{b}}).

\bibitem[{\citenamefont{Liang et~al.}(2014)\citenamefont{Liang, Liu, Mo, Liu,
  Tian, Zhou, Zhang, Zhou, Hou, and Peng}}]{liang2014influence}
\bibinfo{author}{\bibfnamefont{Y.-C.} \bibnamefont{Liang}},
  \bibinfo{author}{\bibfnamefont{R.-S.} \bibnamefont{Liu}},
  \bibinfo{author}{\bibfnamefont{Y.-F.} \bibnamefont{Mo}},
  \bibinfo{author}{\bibfnamefont{H.-R.} \bibnamefont{Liu}},
  \bibinfo{author}{\bibfnamefont{Z.-A.} \bibnamefont{Tian}},
  \bibinfo{author}{\bibfnamefont{Q.-y.} \bibnamefont{Zhou}},
  \bibinfo{author}{\bibfnamefont{H.-T.} \bibnamefont{Zhang}},
  \bibinfo{author}{\bibfnamefont{L.-L.} \bibnamefont{Zhou}},
  \bibinfo{author}{\bibfnamefont{Z.-Y.} \bibnamefont{Hou}}, \bibnamefont{and}
  \bibinfo{author}{\bibfnamefont{P.}~\bibnamefont{Peng}}, \bibinfo{journal}{J.
  Alloy. Comp.} \textbf{\bibinfo{volume}{597}}, \bibinfo{pages}{269}
  (\bibinfo{year}{2014}).

\end{thebibliography}

\end{document}